\documentclass{cup-hpl}

\begin{document}

\newtheorem{theorem}{Theorem}

\shorttitle{Generation of polarized particle beams} 

\shortauthor{M.B\"uscher et al.}

\title{Generation of polarized particle beams at relativistic laser intensities}

\author[1,2]{Markus B\"uscher\corresp{Peter Gr\"unberg Institut (PGI-6), Forschungszentrum J\"ulich, Wilhelm-Johnen-Str. 1, 52425 J\"ulich, Germany, \email{m.buescher@
fz-juelich.de}}}
\author[1,2]{Anna H\"utzen}
\author[3,4]{Liangliang Ji\corresp{Shanghai Institute of Optics and Fine Mechanics, Chinese Academy of Sciences, 201800 Shanghai, China, \email{jill@siom.ac.cn}}}
\author[5,6]{Andreas Lehrach}

\address[1]{Peter Gr\"unberg Institut (PGI-6), Forschungszentrum J\"ulich, Wilhelm-Johnen-Str. 1, 52425 J\"ulich, Germany}
\address[2]{Institut f\"ur Laser- und Plasmaphysik, Heinrich-Heine-Universit\"at D\"usseldorf, Universit\"atsstr. 1, 40225 D\"usseldorf, Germany}
\address[3]{State Key Laboratory of High Field Laser Physics, Shanghai Institute of Optics and Fine Mechanics, Chinese Academy of Sciences, Shanghai 201800, China}
\address[4]{CAS Center for Excellence in Ultra-intense Laser Science, Shanghai 201800, China}
\address[5]{JARA-FAME (Forces and Matter Experiments), Forschungszentrum J\"ulich and RWTH Aachen University, 52056 Aachen, Germany}
\address[6]{Institut f\"ur Kernphysik (IKP-4), Forschungszentrum J\"ulich, Wilhelm-Johnen-Str. 1, 52425 J\"ulich, Germany}

\begin{abstract}
The acceleration of polarized electrons, positrons, protons and ions in strong laser and plasma fields is a very attractive option to obtain polarized beams in the multi-MeV range. Recently, there has been substantial progress in the understanding of the dominant mechanisms leading to high degrees of polarization, in the numerical modelling of these processes and in their experimental implementation. This review article presents an overview on the current status of the field, on the concepts of polarized laser-plasma accelerators and of the beam polarimetry. 

\end{abstract}


\maketitle

\section{The need for polarized beams}

Spin-polarized particle beams are commonly used in nuclear and particle physics to study the interaction and structure of matter, and to test the Standard Model of particle physics\cite{Moortgat_2005, EDM_2013, G-2_2015, Androic_2018}. In particular, the structure of sub-atomic particles like protons or neutrons is explored to get further insights into QCD\cite{Burkardt_2010} or to probe the nuclear spin structure\cite{COMPASS_2005}. Polarized particle beams are also advantageous to achieve a deeper understanding of nuclear reactions\cite{intro_8}, to search for symmetry violations, to pin down quantum numbers of new particles\cite{EDM_2013, intro_9, ILC_phys, Adlarson2014} or to investigate molecular dynamics\cite{Gay2009, Bederson_2017}.

The technique of producing polarized beams not only depends on the particle species but also on their kinetic energies. For stable ones, such as electrons or protons, polarized sources can be employed with subsequent acceleration in a linear accelerator or a synchrotron. For unstable particles, like muons, polarization-dependent particle decays are exploited\cite{G-2_2015}, while stable secondary beams, like antiprotons, might be polarized in dedicated storage rings by spin-dependent interactions\cite{Antiprotons_2005}. Electron or positron beams also spontaneously polarize in magnetic fields of storage rings due to the emission of spin-flip synchrotron radiation, the so-called Sokolov-Ternov effect\cite{Sokolov1964,Sokolov1971, Mane_2005}. This effect was first experimentally observed with low degrees of polarization\cite{ACO, VEPP} and later utilized at several electron rings to generate a highly polarized beams during storage\cite{SPEAR, ELSA, CBAF, CESR, LEP, HERA, Barber_1993}.

All the above scenarios still rely on conventional particle accelerators that are typically very large in scale and budget\cite{Mane_2005}. In circular accelerators, depolarizing spin resonances must be compensated by applying complex correction techniques to maintain the beam's polarization\cite{ZGS, AGS, KEK, IUCF, SATURNE, RHIC, COSY}. In linear accelerators, such a reduction of polarization can be neglected due to the very short interaction time between particle bunches and accelerating fields. 

Concepts based on laser-driven acceleration at extreme light intensities have been promoted during the last decades. Ultra-intense and ultra-short laser pulses can generate accelerating fields in plasmas that are at the order of TV/m, about four orders of magnitudes higher compared to conventional accelerators. The goal, therefore, is to build the next generation of highly compact and cost-effective accelerator facilities using a plasma as the accelerating medium, see e.g.\,Ref.\cite{EUPRAXIA}. However, despite many advances in the understanding of the phenomena leading to particle acceleration in laser-plasma interactions, a largely unexplored issue is how an accelerator for strongly polarized beams can be realized. In simple words, there are two possible scenarios: either the magnetic laser or plasma fields can influence the spin of the accelerated beam particles, or the spins are too inert, such that a short acceleration has no influence on the spin alignment. In the latter case, the polarization would be maintained throughout the whole acceleration process but a pre-polarized target would be required.

In this article, we review the concepts and methods that could lead to generation of polarized particle beams based on ultra-intense lasers. We focus on two main approaches. The first one is devoted to collision between unpolarized high-energy electron beams and ultra-relativistic laser pulses, introduced in Sec.\,\ref{ssec:Polarization build-up from interactions with relativistic laser pulses}. Section \ref{ssec:Polarized beams from pre-polarized targets} focuses on concepts for pre-polarized targets for sequential particle acceleration. Suitable targets are described in Sec.\,\ref{sec:targets}.

\section{Concepts}
\label{sec:concepts}

\subsection{Polarization build-up from interactions with relativistic laser pulses}
\label{ssec:Polarization build-up from interactions with relativistic laser pulses}

Strong-field QED processes --- like nonlinear Compton scattering and radiation reactions --- can strongly modify the dynamics of light charged particles, such as electrons or positrons. Analogous to the Sokolov-Ternov effect in a strong magnetic field, electrons can rapidly spin polarize in ultra-strong laser fields due to an asymmetry in the rate of spin-flip transitions, {\em i.e.}, interactions where the spin changes sign during the emission of a $\gamma$-ray photon. Several such scenarios have been discussed in literature, for a more quantitative discussion we refer to Sec.\,\ref{sec:qedcalculations}:

\begin{enumerate}

\item Li et al.\ describe electron radiative spin effects by a Monte-Carlo spin-resolved radiation approach in the local constant field approximation\cite{Li2019,Guo2020}. Due to a spin-dependent radiation reaction, a monochromatic, elliptically polarized laser pulse can split an initially unpolarized relativistic electron ensemble along the propagation direction into two oppositely transversely polarized parts, see Fig.\,\ref{fig:Li2019_Fig1}.

\begin{figure}[ht]
	\centering
	\includegraphics[scale=0.26]{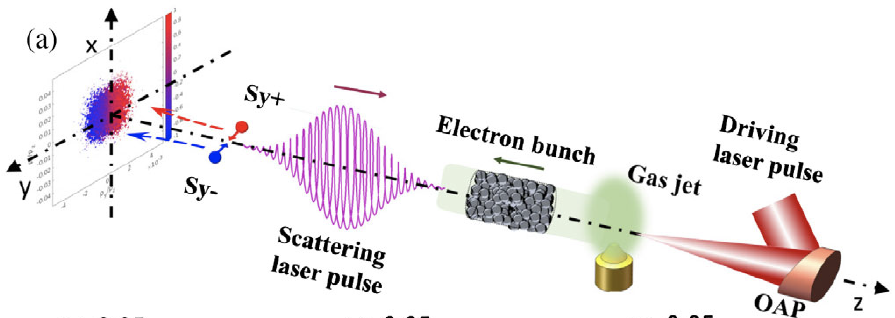}
	\caption{Scenario of generation of spin-polarized electron beams via nonlinear Compton scattering: a relativistic electron bunch generated by laser-wakefield acceleration collides head-on with an elliptically polarized laser pulse and splits along the propagation direction into two parts with opposite transverse polarization\cite{Li2019}.}
	\label{fig:Li2019_Fig1}
\end{figure}

A similar spin-dependent deflection mechanism is found by Geng et al.\cite{Geng2020} who study the spin-correlated radiation-reaction force during the interaction of an initially polarized electron bunch with a linearly polarized laser pulse. The discovered mechanism dominates over the Stern-Gerlach force, which can provide a new perspective to study spin-dependent QED effects.

\item Del Sorbo et al.\cite{DelSorbo2017,Seipt2018} calculate the rate of spin-flip transitions for electrons circulating at the magnetic nodes of two colliding, circularly polarized laser pulses, cf.\,Fig.\,\ref{fig:DelSorbo2018_Fig1}. They find that a sizeable 
($\gtrsim$50\%) spin polarization is expected in one laser period for lasers of intensity within the reach of next-generation laser systems. 

\begin{figure}[ht]
	\centering
	\includegraphics[scale=0.3]{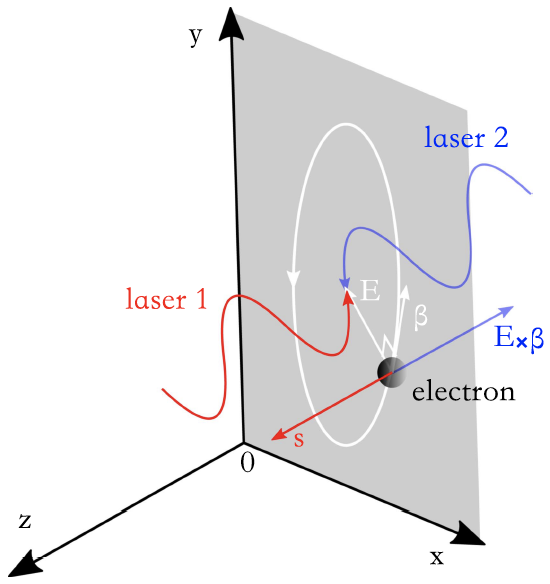}
	\caption{Schematic representation of the electron spin polarization employing the standing wave of two colliding, circularly polarized laser pulses\cite{DelSorbo2018}.}
	\label{fig:DelSorbo2018_Fig1}
\end{figure}

However, Del Sorbo et al.\cite{DelSorbo2018} also demonstrate that the involved electron orbits are unstable and study the robustness of the spin polarization when accounting for the instability of an electron trajectory in a magnetic node using a deterministic model for the radiation-reaction force. They point out that depolarization effects due to chaotic spin precession may strongly limit the achievable electron polarization. In addition, stochasticity --- that may affect the rate of migration of the electrons from the magnetic node --- needs further investigation. For these reasons, a more promising approach seems to be:

\item Radiative polarization of high-energy electron beams in collisions with ultrashort pulsed bichromatic laser fields has been proposed in Refs.\cite{Seipt2019,Song2019}. The scheme is depicted in Fig.\,\ref{fig:Seipt2019_Fig1} and is based on the asymmetric distribution of the field structure that deflects spin-up/down electrons via quantum radiation-reaction. 

\begin{figure}[t]
	\centering
	\includegraphics[scale=0.15]{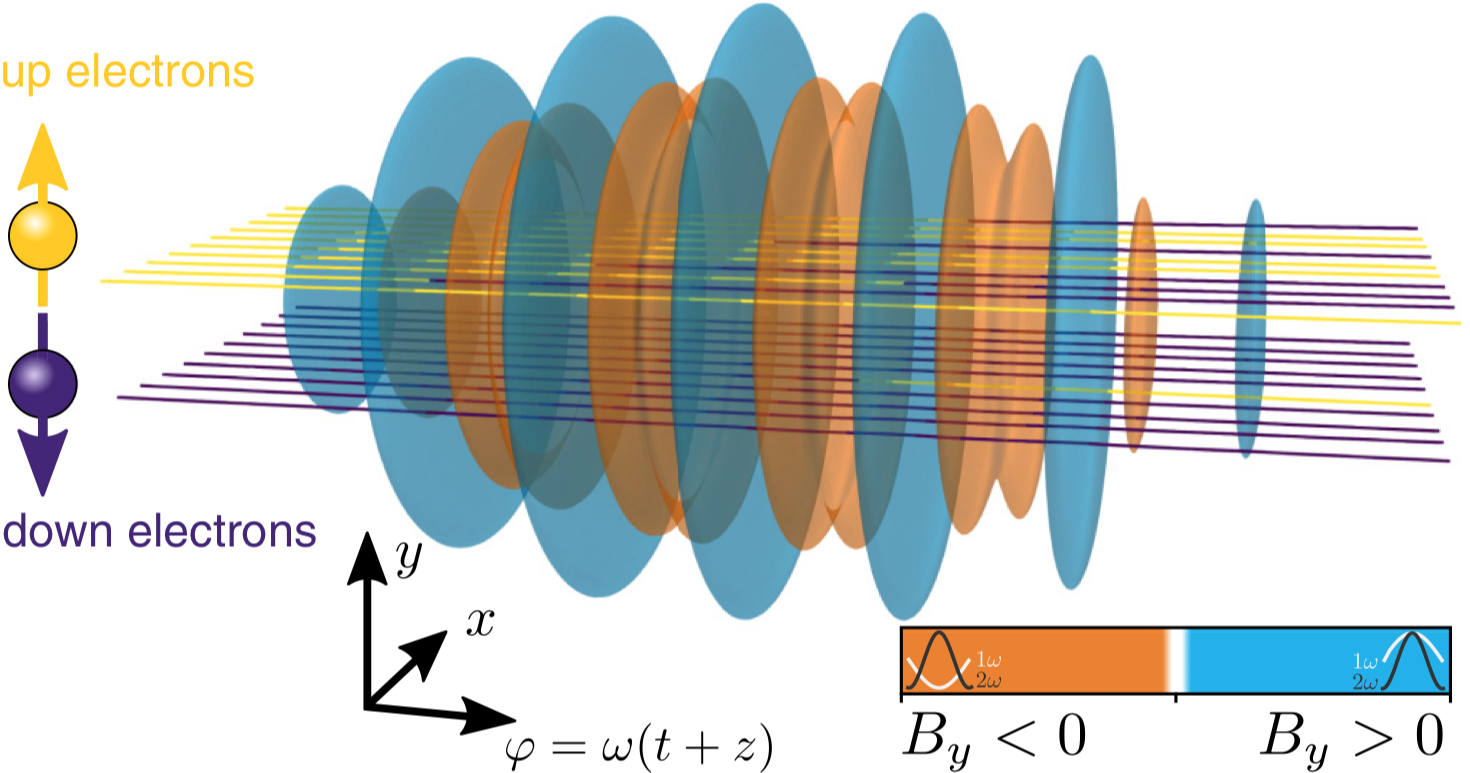}
	\caption{Electrons propagating through a bichromatic laser pulse perform spin flips dominantly in certain phases of the field: electrons initially polarized along the +$y$ direction (yellow trajectories) flip their spin to down (trajectory colored purple) dominantly when $B_\mathrm{y}$\,$>$\,$0$, and this is where 1$\omega$ and 2$\omega$ add constructively (blue contours). The opposite spin flip dominantly happens when $B_\mathrm{y}$\,$<$\,$0$ where the 1$\omega$ and 2$\omega$ components of the laser are out of phase (orange contours)\cite{Seipt2019}.}
	\label{fig:Seipt2019_Fig1}
\end{figure}

\item For the production of polarized positron beams, Chen et al.\cite{Chen2019} employ a similar scheme as used in Refs.\cite{Seipt2019,Song2019} for the electrons. An intense linearly polarized two-color laser pulse collides head-on with an unpolarized relativistic electron beam, resulting in emission of photons in  forward direction which subsequently decay into polarized $e^+$/$e^-$ pairs, with spins parallel and antiparallel to the laser's magnetic field direction, respectively, and with a small divergence angle in the propagation direction (see Fig.\,\ref{fig:Chen2019_Fig1}).

\begin{figure}[ht]
	\centering
	\includegraphics[scale=0.65]{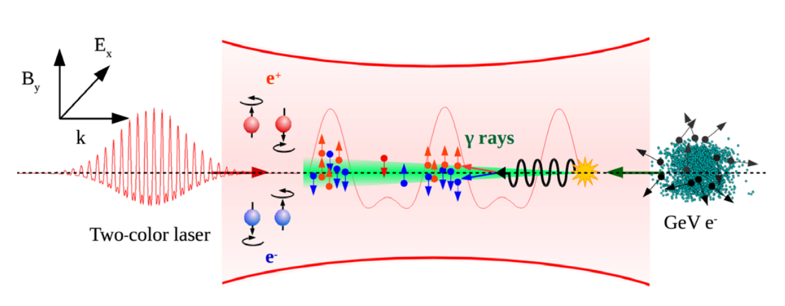}
	\caption{Scheme of laser-based polarized positron beam production\cite{Chen2019}.}
	\label{fig:Chen2019_Fig1}
\end{figure}

Wan et al.\cite{Wan2020} suggest to use an ultra-intense elliptically polarized laser pulse that collides head-on with an unpolarized electron bunch (similar to Ref.\cite{Li2019} for electrons). Again, the radiated high-energy photons decay into polarized electron-positron pairs due to asymmetry of spin-dependent pair-production probabilities. The particles are then split into two beams due to the correlation of the spin-polarization with the particle momenta. In this scheme, the laser field is not asymmetric, and asymmetry of the pair production probability is reflected in the angular separation of the oppositely polarized parts of the beam. This is in contrast to the work of Chen et al.\cite{Chen2019}, where an asymmetric two-color laser field is applied for positron polarization, yielding though considerable less polarization and larger angular spreading. Finally, Li et al.\cite{Li2020} investigate theoretically the feasibility of production of a longitudinally polarized relativistic positron beam via the interaction of a circularly polarized laser pulse with a fully longitudinally spin-polarized counter-propagating relativistic electron beam in the quantum radiation-dominated regime.

\end{enumerate}

\subsection{Polarized beams from pre-polarized targets}
\label{ssec:Polarized beams from pre-polarized targets}

\begin{enumerate}

\item Wen et al.\cite{Wen2019} and Wu et al.\cite{VortexAcceleration2019} have put forward a method for generating intense polarized electron beams. It is based on electron polarization of a gas jet via photo-dissociation by a circularly polarized UV laser pulse followed by electron laser-wakefield acceleration (LWFA) by an intense laser pulse. This scheme is illustrated in Fig.\,\ref{fig:Wu_Fig1a}.

\begin{figure}[ht]
	\centering
	\includegraphics[scale=0.4]{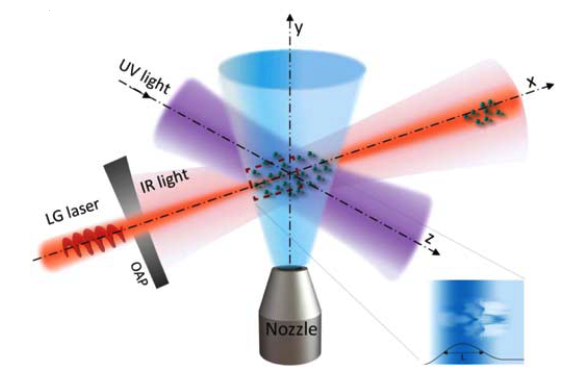}
	\caption{Sketch of the all-optical laser-driven polarized electron acceleration scheme using a pre-polarized target\cite{VortexAcceleration2019}.}
	\label{fig:Wu_Fig1a}
\end{figure}

Assuming that at the moment of irradiation with the accelerating multi-TW laser pulse the electrons in the target are fully polarized (cf.\ Sec.\,\ref{sec:targets} for target details) one has to optimize the injection into the wake field and the subsequent acceleration to multi-MeV energies such that a high degree of electron polarization is maintained. These processes can be modelled with the help of full three-dimensional particle-in-cell (PIC) simulations, incorporating the spin dynamics via the Thomas-Bargmann Michel Telegdi (T-BMT) equation (cf.\,Eq.\,\ref{Eq:TMBT})\cite{Vieira2011,Scaling-laws_2020}. A couple of such codes have been developed recently and used for the modelling of polarized electron\cite{Wen2019,VortexAcceleration2019,BeamDriven2019,SpinFilter2020} and proton\cite{HPLSE2019,IJMP2019,EAAC2020,ProtonBeams2020} beam generation. 

\item In a subsequent paper Wu et al.\cite{BeamDriven2019} apply their scheme developed in Ref.\cite{VortexAcceleration2019} to wakefield acceleration driven by a particle beam (PWFA). In this scheme, the unpolarized electron driver beam can be generated via the well-understood LWFA. The electron-beam driver is free of the prepulse issue associated with a laser driver, thus eliminating possible depolarization of the pre-polarized gas due to ionization by a prepulse.

\item First applications of the pre-polarized targets employed in Refs.\,\cite{Wen2019,VortexAcceleration2019,BeamDriven2019,SpinFilter2020} were actually aiming at the laser-induced acceleration of proton beams\cite{HPLSE2019,IJMP2019,EAAC2020,ProtonBeams2020}. This is because protons have much smaller magnetic moments and, therefore, their spin alignment in the plasma magnetic fields is much more inert as compared to electrons. Also, from the target point-of-view polarized nuclei can be provided more easily than electrons (cf.\ Sec.\,\ref{sec:targets}) and the necessary proton polarimetry can be achieved straight-forward (Sec.\,\ref{sec:polarimetry}). Figure\,\ref{fig:ProtonBeams2020_Fig1} shows the schematic layout of a laser-based accelerator for polarized proton beams, which is simpler than the set-up from Fig.\,\ref{fig:Wu_Fig1a} because the 234.62\,nm UV light for Cl ionization is not required. The first description of this scheme can be found in Ref.\,\cite{HPLSE2019}.

\begin{figure}[ht]
	\centering
	\includegraphics[scale=0.35]{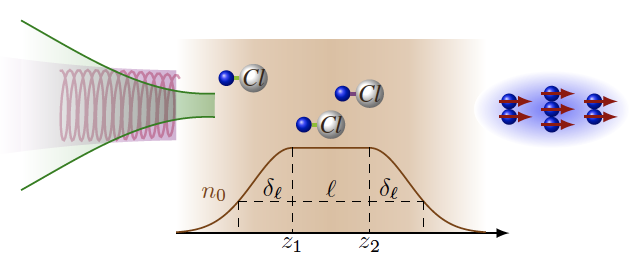}
	\caption{Schematic diagram showing laser acceleration of polarized protons from a dense hydrogen chloride gas target (brown). HCl molecules are initially aligned along the accelerating laser (indicated by the green area) propagation direction via a weak IR laser. Blue and white balls represent nuclei of hydrogen and chlorine atoms, respectively. Before the acceleration, a weak circularly polarized UV laser (purple area) is used to generate the polarized atoms along the longitudinal direction via molecular photo-dissociation. The brown curve indicates the initial density distribution of the gas-jet target. The polarized proton beam is shown on the right (blue) with arrows (red) presenting the polarization direction\cite{ProtonBeams2020}.}
	\label{fig:ProtonBeams2020_Fig1}
\end{figure}

\item The first attempt to experimentally study spin effects during laser-induced acceleration is based on a nuclear polarized $^3$He target\cite{PSTP2015,Unpolarized3He}. These experiments profit from the fact that hyperpolarized $^3$He gas can be produced rather easily and maintains its nuclear polarization over several days at ambient room temperature and small magnetic holding fields (Sec.\,\ref{sec:targets}). The main goal of these studies is to demonstrate nuclear polarization conservation in a (laser-induced) plasma. This would open the possibility of inertial confinement fusion with spin-polarized fuel, in which the cross-sections for nuclear fusion reactions can be enhanced, leading to higher energy yields compared to the case of unpolarized fuel\cite{Polarized Fusion2016}. Another goal is to realize an intense spin-polarized $^3$He-ion source which is extremely challenging with conventional approaches. The main experimental challenge --- besides the preparation of the polarized Helium target --- is the demonstration of laser-induced ion acceleration out of gas-jet targets. This has recently been achieved in a feasibility study with an unpolarized gas-jet target performed at PHELIX, GSI Darmstadt, where Helium ions with energies of a few MeV have been observed, see Fig.\,\ref{fig:Unpolarized3He_Fig6}. Thus, the ion energies are sufficiently high for the polarimetry (see Sec.\,\ref{sec:polarimetry}) in a beam time with polarized target scheduled for fall 2020.

\begin{figure}[ht]
	\centering
	\includegraphics[scale=0.22]{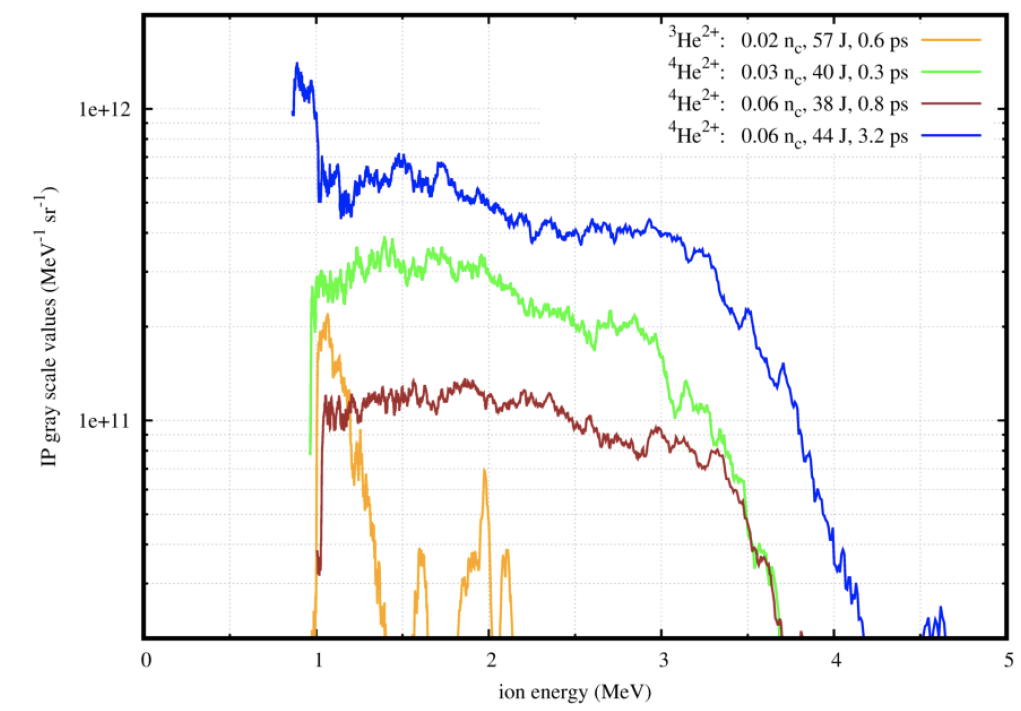}
	\caption{Measured $^{3,4}$He$^{2+}$ energy spectra accelerated from unpolarized Helium gas jets\cite{Unpolarized3He}.}
	\label{fig:Unpolarized3He_Fig6}
\end{figure}

\end{enumerate}

\section{Theoretical background}
\label{sec:theory}

\begin{figure}[b]
	\centering
	\includegraphics[scale=0.27]{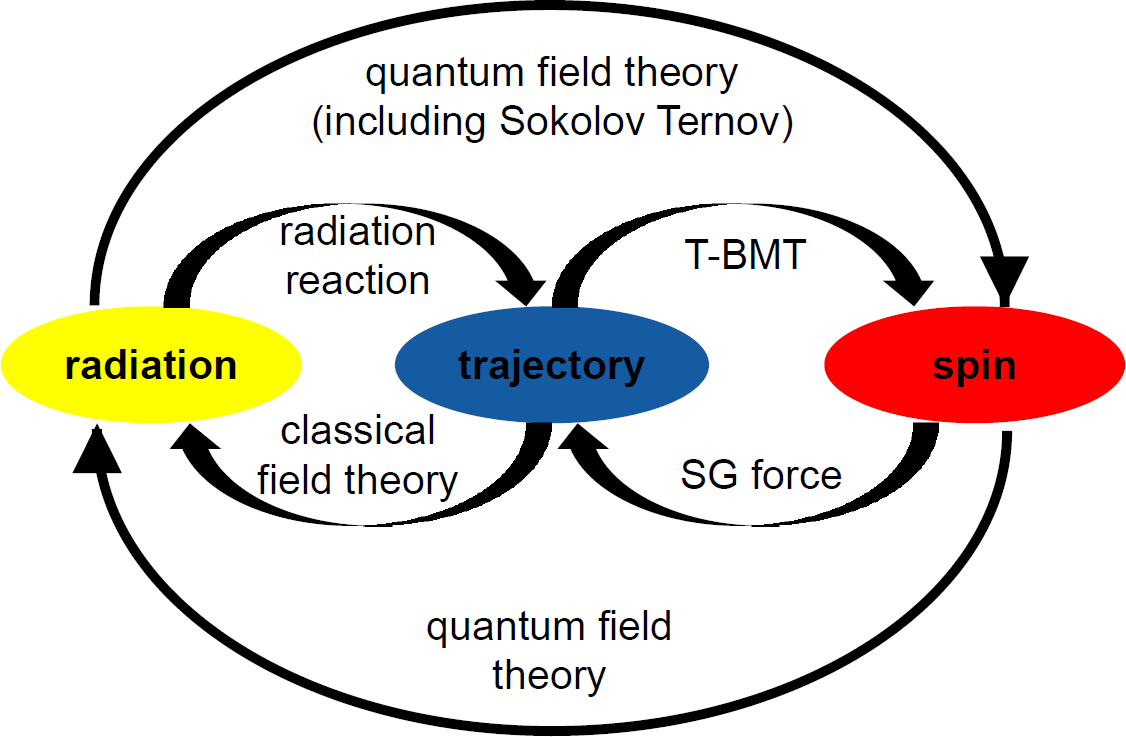}
	\caption[overview]{Sketch of the interplay between single particle trajectories (blue), spins (red) and radiation (yellow)\cite{Scaling-laws_2020}.}
	\label{fig:scheme_johannes}
\end{figure}

It is still an issue of current research how particle spins are affected by the huge electromagnetic fields that are inherently present in laser-induced plasmas or in the laser fields themselves, and what mechanisms may lead to the production of highly polarized beams. Early attempts to describe these processes can be found in Refs.\,\cite{Kotkin2003,Vieira2011}. A schematic overview of the interplay between single particle trajectories (blue), spin (red), and radiation (yellow) is shown in Fig.\,\ref{fig:scheme_johannes}, details can be found in Ref.\,\cite{Scaling-laws_2020}. 

When particle spins are treated in the semiclassical limit, it is the T-BMT equation\cite{T-BMT} that determines the spin precession of individual particles around the local electromagnetic field lines. If particles with mass $m$, charge $q\cdot e$, anomalous magnetic moment $a$ and velocity $\vec{v}$ move in an electromagnetic field $\vec{E}$, $\vec{B}$ with vanishing gradient, their spin vectors $\vec{s}_\mathrm{i}$ precess according to
\begin{align}
		\frac{d\vec{s}_\mathrm{i}}{dt} = - \vec{\Omega} \times\vec{s}_\mathrm{i} \, \, \, . 
		\label{Eq:PugaSpinNorm}
\end{align}	

In cgs units the rotation frequency $\vec{\Omega}$ is given by \cite{Mane_2005}
\begin{equation}
	\vec{\Omega} = \frac{q\cdot e}{mc}\left[\Omega_\mathrm{B} \cdot \vec{B} -\Omega_\mathrm{v}\left(\frac{\vec{v}}{c}\cdot\vec{B}\right)\frac{\vec{v}}{c} -\Omega_\mathrm{E}\frac{\vec{v}}{c}\times\vec{E}\right] \, \, \, ,
	\label{Eq:TMBT}
\end{equation}
where 
\begin{align}
	\Omega_\mathrm{B} = a+\frac{1}{\gamma}, && \Omega_\mathrm{v} = \frac{a \gamma}{\gamma+1}, && \Omega_\mathrm{E} = a +\frac{1}{1+\gamma} \, \, \, . 
\end{align}	
Spin precession is a deterministic process and can be calulated by treating the spin as an intrinsic electron magnetic moment. In the non-QED regime, only a theory, which contains the T-BMT equation, describes the particle and spin motion in electromagnetic ﬁelds in a self-consistent way. 

In the classical and semi-classical limit, the acceleration of charged particles is treated within the framework of classical field theory. This theory also describes the reaction of the particle motion due to radiation, if the particle energy and/or laser field strength is sufficiently high. Introducing spin into electron dynamics leads to a spin-dependent radiation reaction. The radiation power of electrons in different spin states varies such that they feel a stronger radiation-reaction force when the spins are anti-parallel to the local magnetic field in the rest frame of the radiating electron, which can lead to split of electrons with distinctive spin states. 

The Stern-Gerlach force primarily influences the trajectory of a particle. In general, the radiation-reaction force exceeds the Stern-Gerlach force by far if the particles are relativistic (kinetic energies well above 1\,GeV) or even ultra-relativistic (above 1\,TeV) (see also Ref.\,\cite{Geng2020}). However, there are some field configurations that reverse this situation, so that the radiation-reaction force can be neglected compared to the Stern-Gerlach force (see {\em e.g.} Ref.\,\cite{Flood2015}). 

A direct coupling between single particle spins and radiation fields is treated in the context of quantum field theory. Within this theory, the mechanism that describes the spontaneous self-polarization of an accelerated particle ensemble is known as the Sokolov-Ternov effect. The stochastic spin diffusion from photon emission is a non-deterministic process resulting in the rotation of the spin vector in the presence of a magnetic field with the emission of a photon.

The discussion of the generalized Stern-Gerlach force shows that the trajectories of individual particle are perturbed rather by a change of the particle motion induced by the T-BMT equation than by coupling of the spin to the change of the particles' energy or velocity rates, while even small field variations must be taken into account\cite{Scaling-laws_2020}. With regard to a possible polarization build-up through spin-dependent beam split effects, it is found that a TeV electron beam has the best option to be polarized when the plasma is dense enough and the acceleration distance (time) is large enough. For protons, we do not see any realistic case to build-up a polarization by beam separation. In conventional circular accelerators, the Sokolov-Ternov restores the alignment of the spins in experimentally proven polarization times in the range of minutes or hours, depending on the energy of the beam and the bending radius of the beam in bending magnets. The scaling laws for laser-plasma fields predict that the spins of electron moving in strong ($\approx$\,10$^{17}$\,V/m) fields should be polarized in less than a femtosecond\cite{Scaling-laws_2020}. 

Del Sorbo et al.\cite{DelSorbo2017,DelSorbo2018} proposed that this analog of the Sokolov-Ternov effect could occur in the strong electromagnetic fields of ultra-high-intensity lasers, which would result in a buildup of spin polarization in femtoseconds for laser intensities exceeding $5 \times 10^{22}$\,W/cm$^2$. In a subsequent paper\cite{Seipt2018} they develop a local constant crossed-field approximation of the polarization density matrix to investigate numerically the scattering of high-energy
electrons from short intense laser pulses.

Description of the spin-dependent dynamics and radiation in optical laser fields requires a classical spin vector that precesses during photon emission events following the T-BMT equation. This is accomplished by projecting the spin states after each emission onto a quantization axis. The latter could be the local magnetic field in the rest frame of the radiating electron\cite{Li2019,Chen2019}. Alternatively, Seipt et al.\cite{Seipt2019} suggest that the spin orientation either flips or stays the same, depending on the radiation probability. Recently, it has been pointed out by Geng et al.\cite{Geng_2020a} that by generalizing the Sokolov-Ternov effect, the polarization vector consisting of the full spin information can be obtained.

\section{Model calculations I: Strong field QED}
\label{sec:qedcalculations}

\begin{figure}[t]
	\centering
	\includegraphics[scale=0.33]{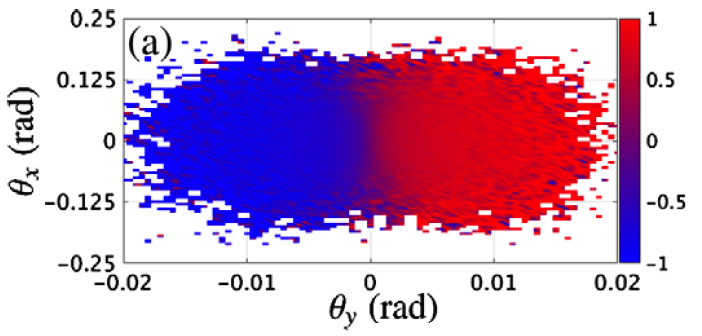}
	\includegraphics[scale=0.33]{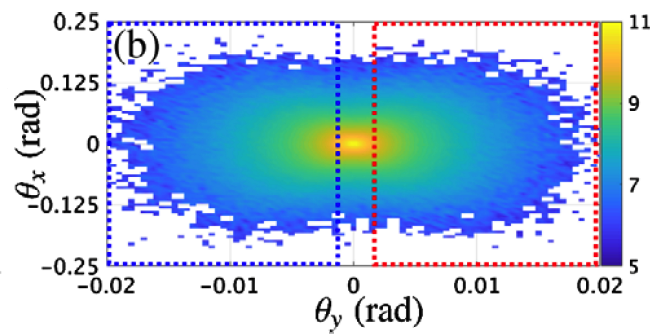}
	\caption{Transverse distribution of the electron spin component $S_{\mathrm{y}}$ as function of the deflection angles $\theta_{\mathrm{x,y}}$ (a). Corresponding logarithmic electron-density distribution (b). The assumed laser peak intensity is $I$\,$\approx$\,1.38$\times$10$^{22}$\,W/cm$^2$ ($a_\mathrm{0}$\,=\,100), wavelength $\lambda$\,=\,1\,$\mu$m, the pulse duration amounts to 5 laser periods, focal radius 5\,$\mu$m, and ellipticity 0.05. The electron bunch with kinetic energy of 4\,GeV and energy spread 6\% has an initial angular divergence of 0.3\,mrad\cite{Li2019}.}
	\label{fig:Li2019_Fig2}
\end{figure}

Figure \ref{fig:Li2019_Fig2} shows the prediction from Li et al.\cite{Li2019} for the splitting of an well-collimated, initially unpolarized electron beam after the interaction with an elliptically polarized laser pulse. It is seen that typical electron deflection angles of a few mrad can be achieved. It is concluded that from a separation of the electron distribution with $\theta_{\mathrm{y}}$\,$>$\,$0$ (or $<$\,0) one can obtain an electron beam with positive (or negative) transverse polarization of roughly 34\%. This number can even be increased to approx.\,70\% by excluding the electrons near $\theta_{\mathrm{y}}$\,=\,$0$. However, this is at the expense of a significantly reduced electron flux and would require a very precise control of the shot-to-shot electron divergence angle.

In a follow-up paper, Guo et al.\cite{Guo2020} investigate stochasticity effects in radiative polarization of a relativistic electron beam head-on colliding with an ultra-strong laser pulse in the quantum radiation-reaction regime. These enhance the splitting effect into the two oppositely polarized parts as described in Ref.\,\cite{Li2019}. Consequently, an increase of the achievable electron polarization by roughly a factor of two is predicted at the same required high accuracy for the selection of the electron deflection angles. 

Another paper from Li et al.\cite{Li2019-Polarimetry} investigates impacts of spin polarization of an electron beam head-on colliding with a strong laser pulse on the emitted photon spectra and electron dynamics in the quantum radiation regime. Using a similar formalism as in Ref.\,\cite{Li2019} they develop an alternative method of electron polarimetry based on nonlinear Compton scattering in the quantum radiation regime. The beam polarization can be measured via the angular asymmetry of the high-energy photon spectrum in a single-shot interaction of the electron beam with a strong laser pulse.

\begin{figure}[t]
	\centering
	\includegraphics[scale=0.16]{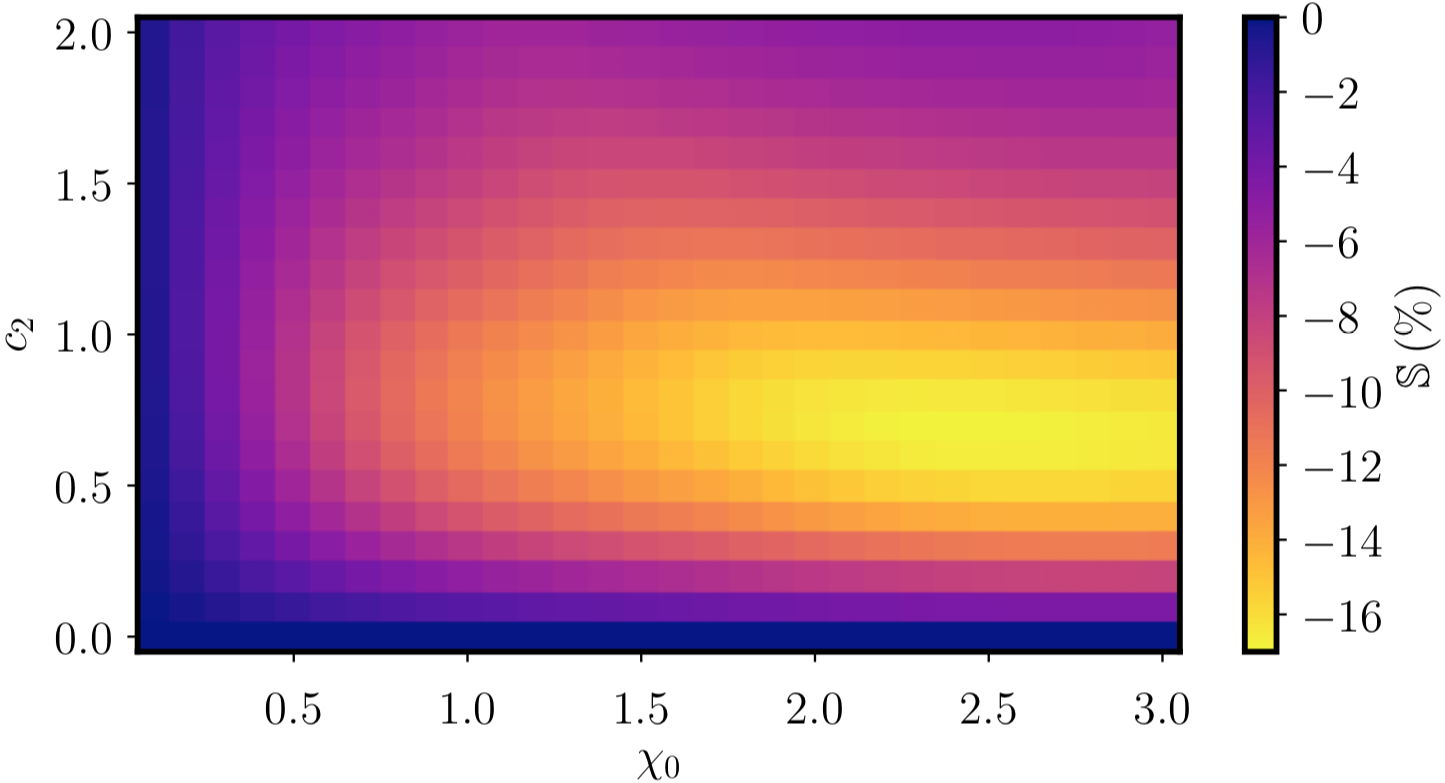}
	\caption{Achievable degree of electron polarization as a function of a quantum nonlinearity parameter $\chi_\mathrm{0}$ and the bichromaticity parameter $c_\mathrm{2}$ (defining the fraction of the total pulse energy in the second harmonic, $c_\mathrm{2}^2/(1 + c_\mathrm{2}^2)$). The calculations have been performed for 5\,GeV electrons colliding with a 161\,fs laser pulse, {\em i.e.}, $a_\mathrm{0}(\chi_\mathrm{0}$\,=\,1)\,=\,16.5\cite{Seipt2019}.}
	\label{fig:Seipt2019_Fig3}
\end{figure}

Seipt et al.\ propose the use of bichromatic laser fields to polarize electron beams and predict a measurable modification of the resulting quantum radiation reaction\cite{Seipt2019}. They describe spin-dependent radiation-reaction effects and use a Boltzmann equation for distribution functions of spin-polarized electrons. They also apply a quasi-classical tracking approach where electrons are pushed classically between photon emissions, and the emissions are treated fully quantum mechanically using a Monte Carlo algorithm employing spin-dependent photon emission rates. In doing so, they can determine optimum parameters for achieving maximum radiative polarization. The $\chi_\mathrm{0}$-$c_\mathrm{2}$ parameter scan shown in Fig.\,\ref{fig:Seipt2019_Fig3} yields a maximum degree of polarization of about 17\%.

\begin{figure}[t]
	\centering
	\includegraphics[scale=0.45]{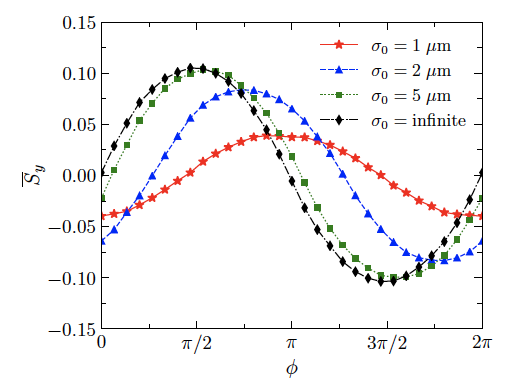}
    \caption{Average polarization $S_{\mathrm{y}}$ as a function of the relative phase $\phi$ of the two-color laser pulse for different laser waist radii $\sigma_{\mathrm{0}}$. The assumed laser intensities are $a_{\mathrm{0,1}}$\,=\,$2a_{\mathrm{0,2}}$\,=\,100, $I_{\mathrm{1}}$\,=\,4$I_{\mathrm{2}}$\,=\,1.37 $\times $10$^{22}$\,W/cm$^{2}$\cite{Song2019}.}
	\label{fig:Song2019_Fig5}
\end{figure}

Song et al.\ find that the electron polarization strongly depends on the relative phase of the two-color laser pulse, see Fig.\,\ref{fig:Song2019_Fig5}. They conclude that with realistic laser parameters maximum degrees of polarization of roughly 10\% seem within reach\cite{Song2019}.

For positrons rather high degrees of polarization seem to be achievable, even for currently achievable laser parameters: Chen et al.\ employ a scenario with initial electron energy of 2\,GeV and laser full intensity $a_\mathrm{0}$\,=\,83. It has been showm that highly polarized positron beams with 2\,$\times$\,10$^4$ particles and a polarization degree of 60\% can be obtained within a small angular divergence of $\sim74$\,mrad\cite{Chen2019}. Wan et al.\ find that their optimal parameters include a laser intensity of the order of $10^{22}$\,W/cm$^2$, an ellipticity of the order of 0.03, a laser pulse duration less than about 10 cycles, and an initial electron energy of several GeVs\cite{Wan2020}. This leads to 86\% polarization of the positron beam, with the number of positrons more than 1\% of the initial electrons. However, as for the electron beams in Ref.\,\cite{Li2019}, the emission angles of the two positron beams with opposite polarization differ by only few mrad. Li et al.\ use a peak laser intensity of $I$\,=\,2.75$\times 10^{22}$\,W/cm$^2$ ($a_\mathrm{0}$\,=\,141), a FWHM pulse duration of 5 laser periods, laser wavelength 1\,$\mu$m, and focal radius 5\,$\mu$m. The initial electron kinetic energy is 10\,GeV, the energy spread 6\%, and the angular divergence 0.2\,mrad\cite{Li2020}. In this scenario, a highly polarized (up to 65\%), intense (up to $10^6$/bunch) positron beam can be obtained.

\section{Model calculations II: Particle-in-cell simulations}
\label{sec:pic}

\subsection{Electron acceleration}
Wen et al.\cite{Wen2019} demonstrate that kA polarized electron beams can be produced via laser-wakefield acceleration from a gas target. For this purpose, they implement the electron spin dynamics in a PIC code which they use to investigate electron beam dynamics in self-consistent three-dimensional particle-in-cell simulations. By appropriately choosing the laser and gas parameters, they show that the depolarization of electrons induced by the laser-wakefield acceleration process can be as low as 10\%. In the weakly nonlinear wakefield regime, electron beams carrying currents of the order of one kA and retaining the initial electronic polarization of the plasma can be produced. The predicted final electron beam polarization and current amount to (90.6\%, 73.9\%, 53.5\%) and (0.31, 0.59, 0.90) kA for $a_\mathrm{0}$\,=\,(1, 1.1, 1.2), respectively. Wen et al.\ point out that compared to currently available conventional sources of polarized electron beams, the flux is increased by 4 orders of magnitude.

\begin{figure}[ht]
	\centering
	\includegraphics[scale=0.3]{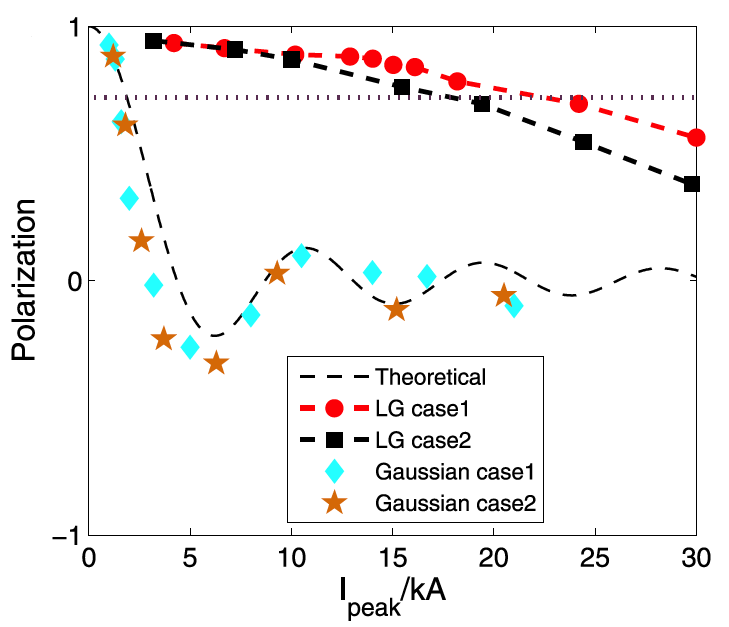}
	\caption{Prediction from Ref.\cite{VortexAcceleration2019} for the achievable electron polarization in dependence of the electron current. More than 80\% polarization can be achieved when a vortex Laguerre-Gaussian (LG) laser pulse is used for the acceleration.}
	\label{fig:VortexAcceleration2019-Fig6}
\end{figure}

Based on similar PIC simulations Wu et al.\cite{VortexAcceleration2019} predict even larger electron beam currents via vortex Laguerre-Gaussian (LG) laser-driven wakefield acceleration, see Fig.\,\ref{fig:VortexAcceleration2019-Fig6}. The topology of the vortex wakefield resolves the depolarization issue of the injected electrons. Their method releases the limit on beam flux for polarized electron acceleration and promises more than an order of magnitude boost in peak flux, as compared to Gaussian beams.

Wu et al.\cite{SpinFilter2020} found that the beam polarization depends on the azimuthal angle in plasma wakefield due to the symmetric bubble field. Accordingly, an X-shaped slit ("spin filter") is proposed to significantly enhance the beam polarization of the accelerated electrons. A beam polarization of about 80\% is achieved by filtering out the low-polarization population using the slit, while the initial polarization is only about 35\%. 

\begin{figure}[ht]
	\centering
	\includegraphics[scale=0.3]{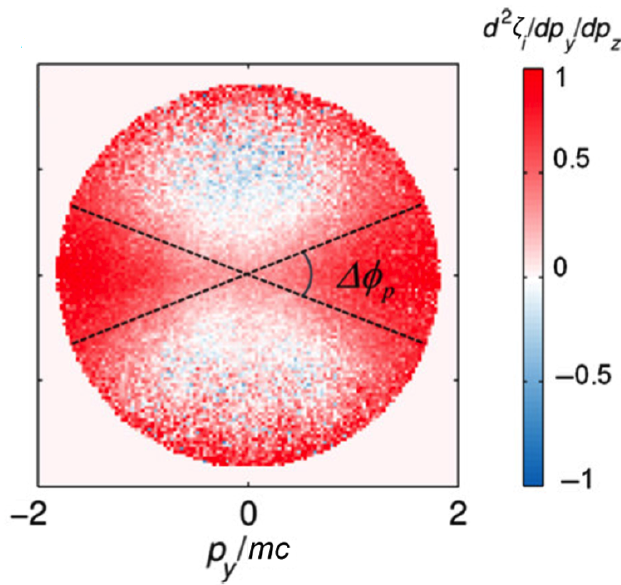}
	\caption{Electron polarization distributions in the transverse phase space during LWFA\cite{SpinFilter2020}.}
	\label{fig:Wu2020_Fig1}
\end{figure}

\subsection{Heavy particles}
\begin{figure}[ht]
	\centering
	\includegraphics[scale=0.21]{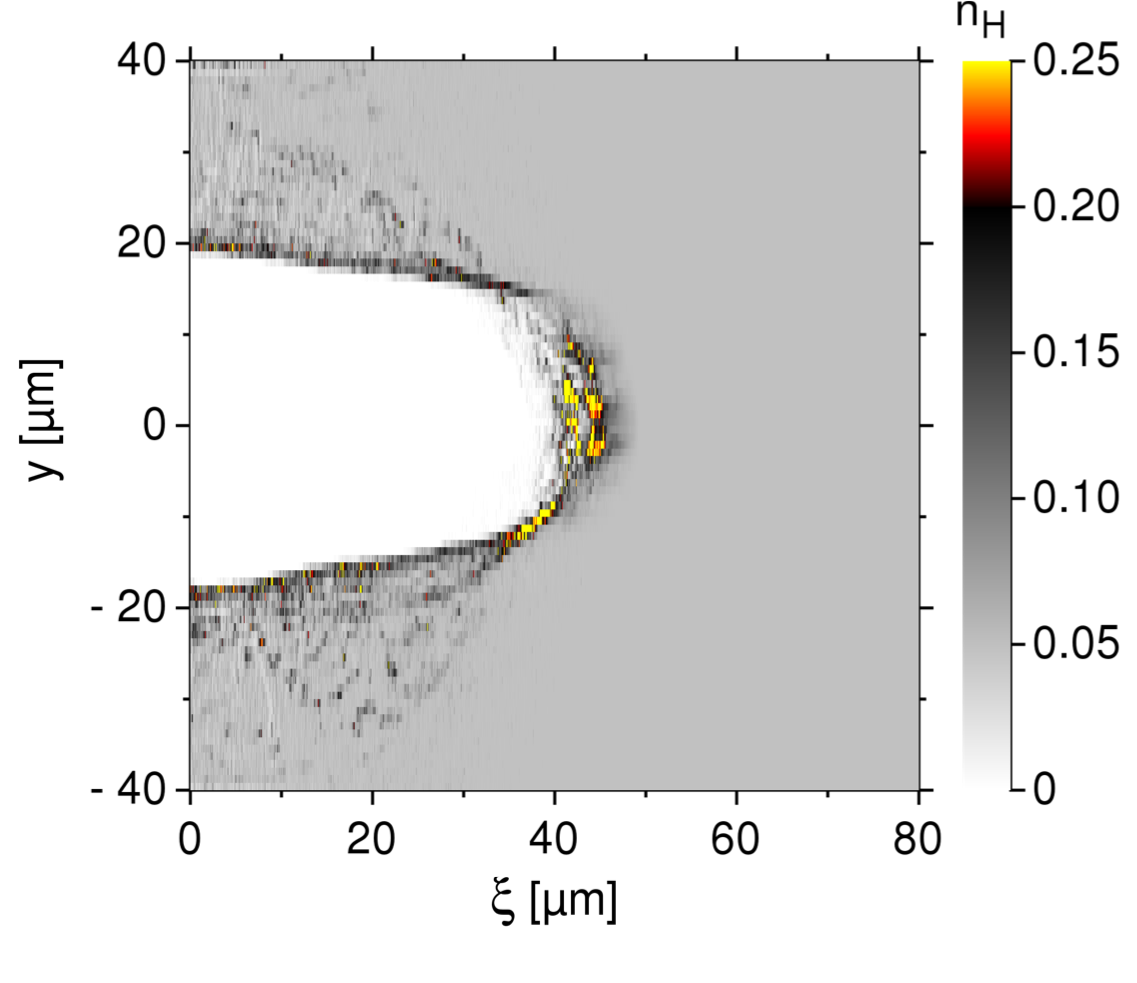}
	\includegraphics[scale=0.21]{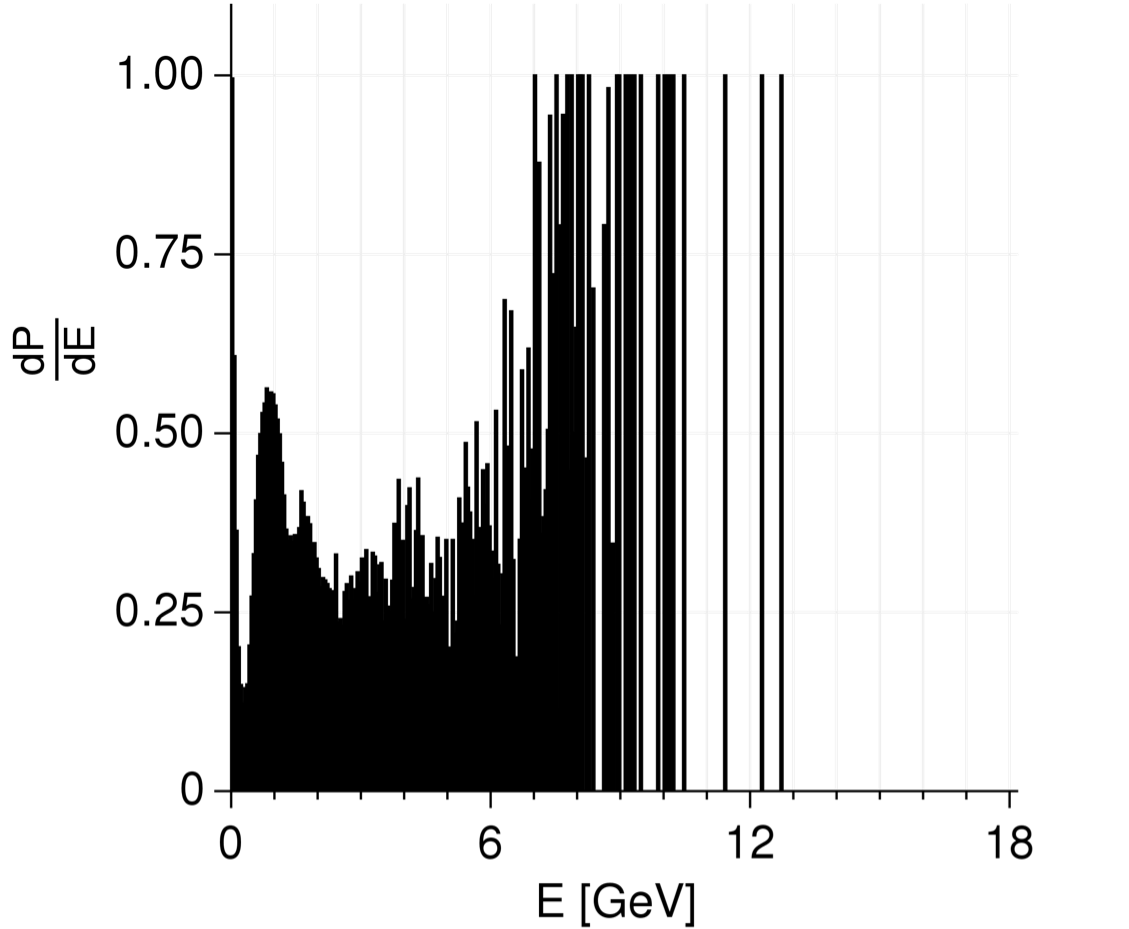}
	\caption{3d PIC simulation of proton acceleration assuming a gaseous HCl target with a hydrogen density of $8.5\,\times\,10^{19}$\,cm$^{-3}$ and a circularly polarized laser pulse with 800\,nm wavelength and a normalized amplitude of $a_\mathrm{0}$\,=\,200. Shown are the simulated proton density (left) and polarization as function of the proton energy (right)\cite{EAAC2020}.}
	\label{fig:EAAC2020_Fig7}
\end{figure}

H\"utzen et al.\cite{HPLSE2019,IJMP2019,EAAC2020} present the first scheme of a laser-based accelerator for polarized particle beams using 3d PIC simulations with explicit spin treatment. They can show that the proton polarization is sufficiently conserved during the acceleration process for foil\cite{HPLSE2019} and gaseous\cite{EAAC2020} targets and, thus, suggest the use of pre-polarized monatomic gases from photo-dissociated hydrogen halide molecules in combination with PW lasers. For an $a_\mathrm{0}$\,=\,200 laser pulse they predict high degrees of polarization at proton energies of a few GeV, see Fig.\,\ref{fig:EAAC2020_Fig7}. Thus, it suggests the use of pre-polarized mono-atomic gases from photo-dissociated hydrogen halide molecules in combination with 10\,PW class lasers.

\begin{figure}[ht]
	\centering
	\includegraphics[scale=0.36]{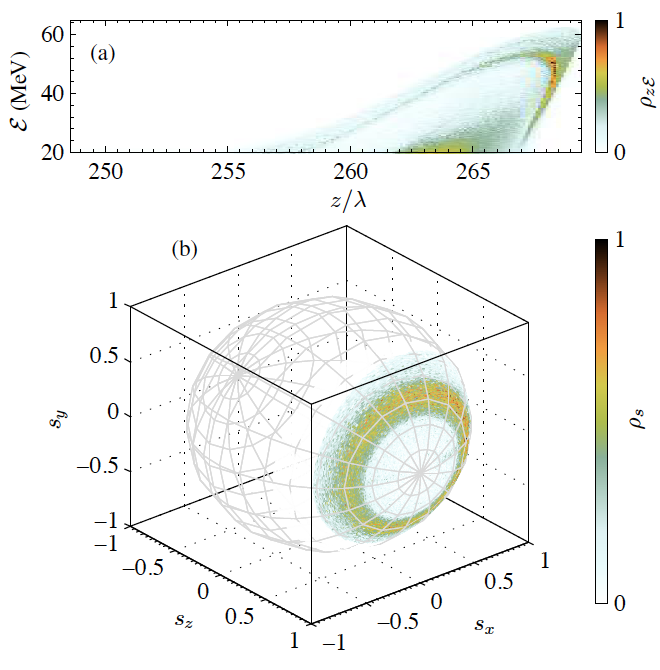}
	\caption{3d PIC simulation for a gaseous HCl target with molecular density of $10^{19}$\,cm$^{-3}$ and 1.3\,PW laser of phase-space distribution (a), and spin spread of protons with energy $E$\,$>$\,20\,MeV on the Bloch sphere (b)\cite{ProtonBeams2020}.}
	\label{fig:ProtonBeams2020_Fig2}
\end{figure}

Jin et al.\cite{ProtonBeams2020} extend the PIC simulations to smaller, currently achievable, laser powers. They find that proton beams with an energy above 50\,MeV and $\sim80$\% polarization can be obtained (see Fig.\ref{fig:ProtonBeams2020_Fig2}) employing the magnetic vortex acceleration mechanism. Such measurements are now being prepared at the SULF facility of SIOM, Shanghai.

\begin{figure}[ht]
	\centering
	\includegraphics[scale=0.24]{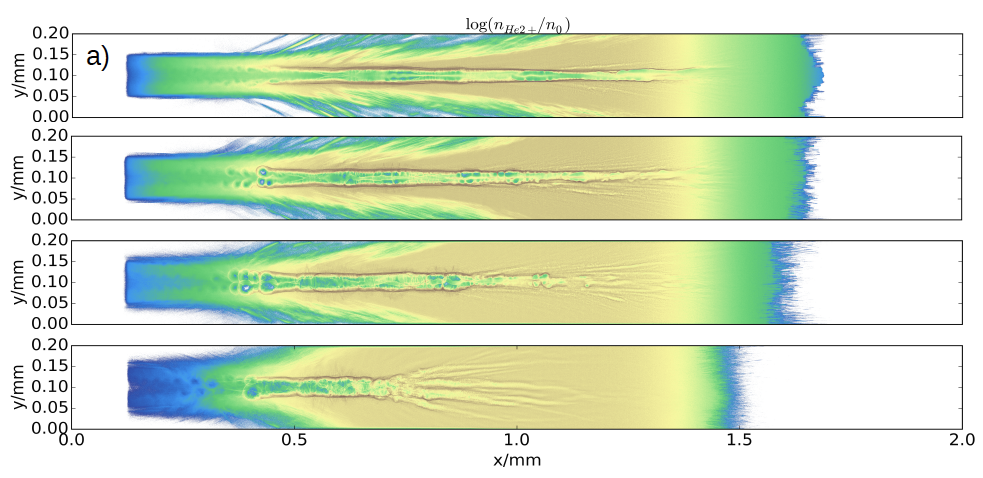}
	\caption{Simulated normalized He$^{2+}$ ion-number density during the passage of a PW laser pulse (6.5\,ps after it entered the simulation box at the left boundary) through an unpolarized Helium gas jet target with 2\%, 3\%, 4\%, 12\% critical density (from top to bottom)\cite{Unpolarized3He}.}
	\label{fig:Unpolarized3He_Fig7}
\end{figure}

No PIC simulations with treatment of spin effects have been carried out for particles heavier than protons yet. So far, only a scan of the parameter space (target density, laser pulse energy and duration) has been published\cite{Unpolarized3He} aiming at the optimization of the ion flux and kinetic energy accelerated in a polarized gas-jet target, see Fig.\,\ref{fig:Unpolarized3He_Fig7}. It is seen that a channel in ion density is generated via a combination of strong self-focusing and radial ponderomotive expulsion of electrons within the first 0.5\,mm of the gas target, followed by filamentation and hosing for larger times. In general, a cleaner and longer channel is generated at lower densities; whereas the laser pulse is prone to filamentation and radial dispersion with increasing density. The influence of these structures on the ion spins is subject of ongoing simulations in the framework of \textit{Ju}SPARC\cite{JuSPARC}.

\section{Lessons learned from theoretical studies}

From the literature outlined in Secs.\,\ref{sec:concepts}--\ref{sec:pic} it becomes clear that a wealth of (mostly theoretical) pathways towards the realization of laser-induced polarized particle acceleration have been put forward in recent years. These concepts strongly differ for the various particle species. In some cases it is necessary to wait for a significant progress in laser technology. Our conclusions for a strategy aiming at the speedy realization of laser-induced polarized particle acceleration are:

\begin{enumerate}

\item For currently realistic laser parameters, pre-polarized targets are needed to achieve electron beams with polarizations well above 10\%. Such targets should provide high degrees of electronic polarization ($>$\,50\%) and should allow for operation at laser facilities ({\em e.g.} robustness against electromagnetic pulses and target heating).

\item Due to their three orders smaller magnetic moments, measurable polarization for heavier particles (protons, ions) can only be achieved with nuclear pre-polarized targets.

\item For positrons, no pre-polarized targets can be realized. Here, high degrees of polarization (90\%) can be obtained from the scattering of PW laser pulses off an unpolarized relativistic electron beam (which can be laser-generated). Such schemes require precise control of all involved beam pointings (to the few-mrad level).

\item Gas-jet targets are preferable to foil targets since they allow the operation with state-of-the-art kHz laser systems. Low-density targets are also less challenging in terms of depolarizing effects. 

\end{enumerate}

\section{Experimental techniques I: Polarized targets}
\label{sec:targets}

For the experimental realization of polarized beam generation from laser-induced plasmas, the choice of the target is a crucial point. Pre-polarized solid foil targets suitable for laser acceleration via target normal sheath acceleration (TNSA) or radiation-pressure acceleration (RPA) are not available yet and their realization seems extremely challenging. In previous experiments hydrogen nuclear polarization has mostly been realized through a static polarization, {\em e.g.}, in frozen spin targets\cite{Keith2012} or with polarized $^{3}$He gas. For proton acceleration only polarized atomic beam sources based on the Stern–Gerlach principle are available until now which, however, offer a too small particle density\cite{Nass2003}. To laser-accelerate polarized electrons and protons, a new approach with dynamically polarized hydrogen gas targets is needed. A statically polarized $^{3}$He target as well as a dynamically polarized hydrogen target for protons as well as a hyperpolarized cryogenic target for the production and storage of polarized H$_{2}$, D$_{2}$ and HD foils are being prepared at the Forschungszentrum J\"ulich within the ATHENA project.

\subsection{Static polarization: $^3$He}
\begin{figure}[t]
	\centering
	\includegraphics[scale=0.6]{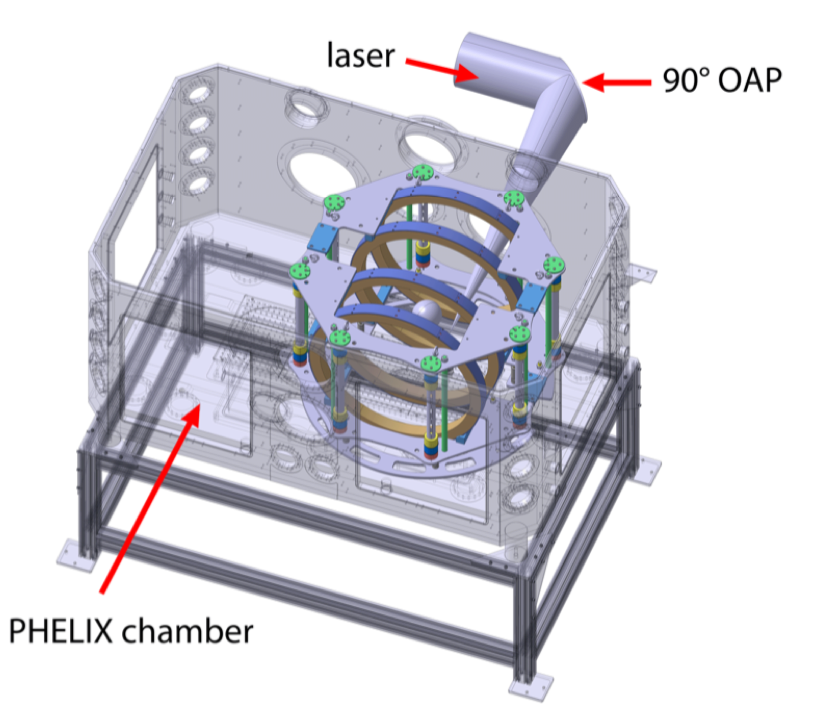}
	\caption{Perspective view of the 3D model of the fully mounted magnetic system inside the PHELIX chamber\cite{3HeTarget2016,Engin2015}.}
	\label{fig:3HeTarget_Fig8}
\end{figure}

In order to develop a laser-driven spin-polarized $^{3}$He-ion beam source available for nuclear-physics experiments as well as for the investigation of polarized nuclear fusion, one challenge is the provision of a properly statically polarized $^{3}$He gas-jet target. The essential components of such a target are: a magnetic holding field for storing pre-polarized $^{3}$He gas for long time duration within the PHELIX target chamber and a non-magnetic nozzle for providing the desired gas-jet target (cf.\,Fig.\,\ref{fig:3HeTarget_Fig8})\cite{Engin2015}. All components must be designed such that the polarization is maintained sufficiently long for the experiments. A relaxation time of  20.9\,h could already be achieved for a prototype of the setup\cite{IEEE2016}.

The magnetic holding field consists of an outer Halbach array composed of an upper and lower ring of 48 NdFeB permanent magnets with 1100\,mm in diameter together with an inner Helmholtz-coil array consisting of four single Helmholtz coils. In the Halbach array the permanent magnets are stacked at an optimum distance such that its field homogeneity is sufficiently high to maintain the nuclear $^3$He polarization. The Helmholtz-coils are oriented thus that their magnetic field is aligned parallel to the laser-propagation direction. A single coil consists of a coiled Cu sheet with a width and thickness of 40\,mm. The outer and inner diameters of the naked Cu coil are 789\,mm and 709\,mm. Both inner coils are separated by 285.75\,mm while the two single front/rear coils have a distance of 218.95\,mm. In contrast to electric coils, the used permanent magnets do not need to be cooled in vacuum and they provide a constant field, even in presence of huge EMPs\cite{Engin2015}. 

The second essential component for the layout of a polarized $^{3}$He target is the gas-jet source. The pre-polarized $^{3}$He gas is delivered at an intrinsic pressure of 3\,bar. By using a pressure booster built of non-magnetic materials, the desired final pressure can be reached (up to 30\,bar). To synchronize the gas flux with the incoming laser pulse a home-made non-magnetic valve with piezos has been prepared. In order to generate a broad plateau-like density distribution with sharp density gradients, a super-sonic de Laval nozzle is used.

\subsection{Dynamic polarization: Protons and electrons}

\begin{figure}[hb]
	\centering
	\includegraphics[scale=0.65]{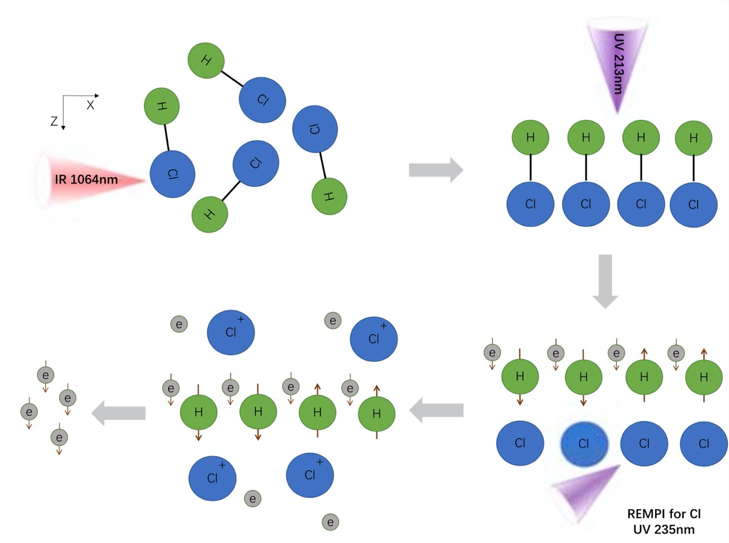}
	\caption{The 1064\,nm IR laser propagates along the $x$ axis to align the bonds of the HCl molecules, then a UV light propagates along the z axis with a wavelength of 213\,nm is used to photo-dissociate the HCl molecules. A 234.62\,nm UV light is used to ionize the Cl atoms. Thermal expansion of the electrons creates large Coulomb field that expels the Cl ions. A fully polarized electron target is therefore produced for sequential acceleration\cite{VortexAcceleration2019}.}
	\label{fig:ElectronTarget}
\end{figure}

For the realization of a dynamically polarized electron and ion source, a novel laser-based target system is under preparation: two laser beams (for proton and three beams for electron polarization) are focused into a gas jet made of bi-atomic linear molecules with at least one Hydrogen atom like, for example, HCl or HBr gas (cf.\,Fig.\,\ref{fig:ElectronTarget}) \cite{HPLSE2019,IJMP2019,JuSPARC,VortexAcceleration2019}. 

\begin{figure}[ht]
	\centering
	\includegraphics[scale=0.5]{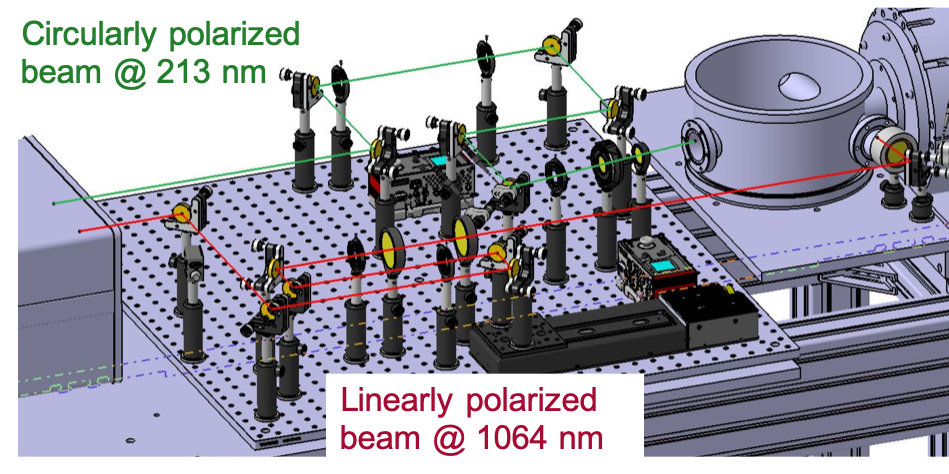}
	\caption{Technical drawing of the optical setup including the JuSPARC\_MIRA laser system and the target chamber of the polarized proton target\cite{JuSPARC}.}
	\label{fig:HClTarget}
\end{figure}

The special feature of the used EKSPLA SL330 series JuSPARC\_MIRA system \cite{JuSPARC} is the simultaneous output of the fundamental wavelength at 1064\,nm and the fifth harmonic at 213\,nm provided by a Nd:YAG crystal serving as active medium. Operating at a repetition rate of 5\,Hz and a pulse duration of 170\,ps the linear polarized fundamental beam is focused onto the gas jet with a pulse energy of 100\,mJ. Thereby, the molecular electric dipole moment $\mu$ is aligned relative to the electric field of the laser light leading to an increased polarization signal. Simultaneously, but under an angle of 90$^\circ$, the strongly focused circularly polarized fifth harmonic beam with an intensity of about 10$^{12}$\,Wcm$^{-2}$ is also guided into the vacuum chamber (cf.\,Fig.\,\ref{fig:HClTarget}). The interaction with the already aligned HCl or HBr molecules leads to a photo-dissociation process by UV excitation and, finally, the polarization of the H nuclei via hyperfine spin beating with a period of about 350\,ps. For the realization of a polarized electron target, unlike for a polarized proton target, an additional third laser at 234.62\,nm UV light which ionize the Cl or Br atoms is needed. The resulting thermal expansion of the electrons creates a large Coulomb field which expels the Cl or Br ions together with their unpolarized electrons\cite{VortexAcceleration2019}.

The fifth harmonic beam is guided by customized optics with highest possible light reflectance (reflectively $>$\,98\,\% at 45$^\circ$ incidence angle provided by \textit{LAYERTEC GmbH}) having a diameter of one inch for a beam diameter of 12\,mm. A quartz quarter-wave plate with two-sided anti-reflection coating from \textit{EKSMA Optics} converts the initially linearly polarized laser beam to circular polarization. Finally, the UV beam is focused below the HCl or HBr nozzle inside the interaction chamber. The fundamental beam at 1064\,nm is guided by standard mirrors with dielectric Nd:YAG coatings and focused to an intensity of about 5$\cdot$10$^{13}$\,Wcm$^{-2}$ into the HCl or HBr gas. The gas is injected into the interaction chamber by a high-speed short-pulse piezo valve that can be operated at maximum 5\,bar inlet-gas pressure to produce a gas density in the range of about 10$^{19}$\,cm$^{-3}$ \cite{JuSPARC, HPLSE2019}. The valve is adjustable in height so that a sufficient amounts of HCl or HBr molecules, which are spread in a cone-like shape, interact with the laser beams by keeping the backing pressure low and thus the molecules' mean free path large enough.

\subsection{Hyperpolarized cryogenic targets}
\begin{figure}[ht]
	\centering
	\includegraphics[scale=0.25]{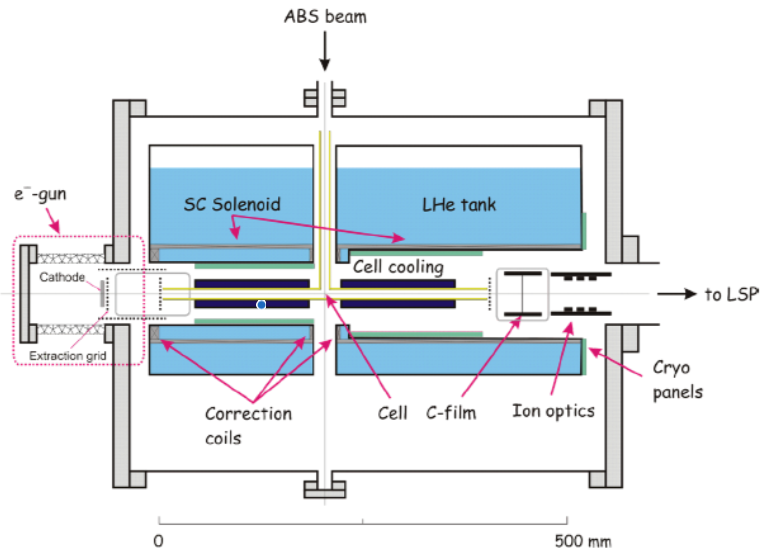}
	\caption{Schematic view of the interaction chamber for production and storage of polarized H$_{2}$, D$_{2}$, HD and HD$_{2}^{+}$ foils\cite{HDmoleculesSPIN2018}.}
	\label{fig:HDmolecules}
\end{figure}

The investigation of the recombination of nuclear polarized hydrogen and deuterium atoms into polarized molecules gives new insights into different fields in physics and chemistry aiming for the optimization of storage-cell gas targets for coming accelerators experiments and the production and handling of polarized fuel for future fusion reactors. In a joined collaboration of the St.\,Petersburg Nuclear Physics Institute (PNPI), the Institute for Nuclear Physics of the University of Cologne and the Institute for Nuclear Physics of the Forschungszentrum J\"ulich the recombination processes on different surfaces, the polarization losses due to wall collisions, and the polarization lifetime of the molecules has been studied. The dedicated experimental setup of the hyperpolarized  cryogenic target is shown in Fig.\,\ref{fig:HDmolecules}. A beam of hydrogen or/and deuterium atoms with selected nuclear and electron spin orientations is produced in a polarized atomic beam source (ABS). These polarized atoms enter a home-made T-shaped storage cell which is 400\,mm long with a 100\,mm long attachment for the atomic inlet (outer diameter of the tube is 14\,mm). In the cell, the atoms can recombine into molecules on the surface of the cell, which is exchangeable to enable measurements with different inner surface coatings on the fused-quartz wall materials, e.g.\,a gold surface, an additional water surface or a coating of Fomblin oil (Perfluropolyether). 

In a next step, a small pipe including an independent cooling and power supply will be installed on one side of the cell having no direct contact with the cell. In this way, molecules can be generated and pre-cooled in the storage cell before they are frozen in the new pipe. Thus, the molecules in the storage cell can still be ionized and accelerated to measure their polarization. After the atomic flow is stopped, the pipe slowly warms up. In this way, the polarization of the molecules that are frozen before can be measured to compare the polarization values of the just recombined molecules and the one that are frozen as ice. The residual gas is pumped by cryogenic panels below 10$^{-8}$\,mbar without gas load to the cell. Using the superconducting solenoid at a temperature of at 4\,K, a magnetic field in the storage cell of up to 1\,T can be generated. Additionally, it focuses the electron beam, which is produced by an electron gun at energies of a few 100\,eV at the left side of the apparatus. The interaction of the polarized atoms and evaporated molecules with the electron beam results in an ionization process. 
Next, the ionized protons and H$_{2}^{+}$ ions are accelerated by a positive electric potential on the cell of up to 5\,kV to the right side. The nuclear polarization of protons/deuterons or the molecular ions H$_{2}^{+}$/D$_{2}^{+}$ and HD$_{2}^{+}$ are measured with a Lamb-shift polarimeter connected to the right end of the apparatus.

\section{Experimental techniques II: Beam polarimetry}
\label{sec:polarimetry}

In order to experimentally determine the degree of polarization of laser-accelerated particle bunches, dedicated polarimeters must be used. Similar devices are widely used in particle physics, for example to determine beam polarizations at classical accelerators. They are typically based on a scattering process with known analyzing power, which converts the information about the beam polarization into a measurable azimuthal angular asymmetry. However, for the case of laser-accelerated particles a couple of peculiar requirements have to be taken into account:

\begin{enumerate}

\item  Due to the time structure of the laser pulses all scattered particles hit the detector within a few 10\,fs. Thus, it must be virtually dead-time free or, more realistically; all particle signals from one laser shot must be integrated up.

\item  The detectors must have a high EMP robustness. This is especially challenging for electronic detectors with on-line readout. 

\item  A high angular resolution is required in some cases, see {\em e.g.} Fig.\,\ref{fig:Li2019_Fig2}.

\item Depending on the phase-space densities of the accelerated particles, it may be required to measure small particle numbers (per laser shot), see {\em e.g.} Ref.\cite{HHUD2014}.

\end{enumerate}

\subsection{Proton and ion polarimetry}

Raab et al.\cite{HHUD2014} report a first polarization measurement of laser-accelerated particles. They developed a proton polarimeter based on the spin dependence of hadronic scattering off nuclei in a silicon foil. These investigations were carried out with protons from unpolarized foil targets, illuminated by 100\,TW accelerating laser pulses of the Arcturus laser facility at D\"usseldorf University. A careful analysis of the measured proton scattering distributions, utilizing analyzing powers from literature, allows one to measure proton-beam polarization with uncertainties as small as approx.\ 10\%. 

\begin{figure}[t]
	\centering
	\includegraphics[scale=0.21]{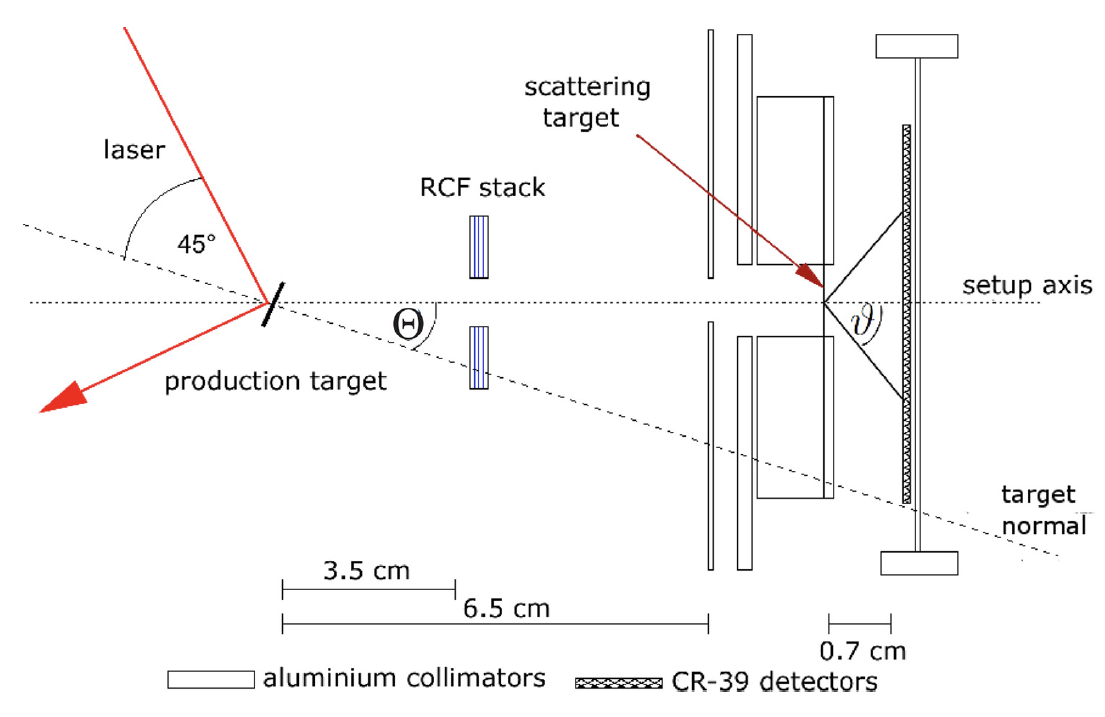}
	\caption{Schematic view of the setup for the proton polarization measurements from Ref.\cite{HHUD2014}. Protons are accelerated from an unpolarized gold foil to energies of about 3\,MeV, are scattered in a silicon foil ("scattering target") and finally detected with CR-39 detectors.)}
	\label{fig:HHUD2014_Fig2}
\end{figure}

For the polarimetry of protons with higher kinetic energies, CH$_2$ (polypropylene foils) instead of Silicon is the proper material for the polarimeter. A new proton polarimeter is now being commissioned and calibrated with polarized protons at COSY-J\"ulich, where beam energies from 45\,MeV up to 2.88\,GeV are available.

\subsection{Electron polarimetry}
\label{Electron polarimetry}

Depending on the electron beam energy, which determines the analyzing power as well as the experimental access to the scattering products, one of the following spin-dependent QED processes can be used for electron polarimetry~\cite{Aulenbacher:2018weg}:
\begin{enumerate}
	\item Mott scattering~\cite{mott, Kessler:1969, Gay:1992}, {\em i.e.\/} scattering off the nuclei in a target, used for beams between 10\,keV and 1\,MeV, often for polarimetry of electron sources at large accelerators

	\item Bremsstrahlung emission in a target~\cite{Olsen:1959zz}, used from about 10\,MeV to a few 100\,MeV, relies on measuring the degree of circular polarization of photons generated when passing the beam through a thin target~\cite{Alexander:2009nb}. Statistical significance of the order of 10\% can be achieved.

	\item M\o ller (or for positron beams Bhabha) scattering~\cite{PhysRev.122.536}, {\em i.e.\/} scattering off the electrons in a target, used from a few 100\,MeV to GeV energies in fixed target experiments at SLAC~\cite{Cooper:1975cu, Band:1997ee, Steiner:1998gf, Anthony:2005pm}, JLab~\cite{Baltzell:2020nvm}, but also at ELSA~\cite{Speckner:2004kt}, MAMI~\cite{Wagner:1990sn}. Precisions down to 0.5\% can be reached~\cite{Hauger:1999iv}.

	\item Compton scattering~\cite{fano}, {\em i.e.\/} scattering off a laser, used for GeV and higher energies, offers high analysing power $\mathcal{O}(1)$, large and precisely known cross-section~\cite{Swartz:1997im} and robust control over experimental systematics. Long-established for measuring longitudinal and transverse polarization, {\em e.g.\/} at SLC~\cite{Abe:2000dq}, LEP~\cite{Placidi:1988nj}, HERA~\cite{Barber_1993, Barber:1994ew}, ELSA~\cite{Doll:1998kp}, MAMI~\cite{Lee:2008ad}, JLab~\cite{Narayan:2015aua}, it is also the method of choice for future colliders~\cite{Adolphsen:2013kya, Boogert:2009ir, Aicheler:2012bya}. Precisions from a few percent down to a few permil can be reached.

\end{enumerate}

The short bunch length typical for plasma-accelerated beams is not a problem for any of these methods, rather an advantage. All methods apart from Compton scattering are destructive. Due to typical energies obtained in laser-wakefield experiments, method 2 is the most applicable technique for diagnosing the degree of polarization of such beams. A new polarimeter for measuring polarization of laser-plasma accelerated electrons is being designed and constructed at DESY.

\section{Acknowledgement}
This work has been carried out in the framework of the \textit{Ju}SPARC (J\"ulich Short-Pulse Particle and Radiation Center) project and has been supported by the ATHENA (Accelerator Technology HElmholtz iNfrAstructure) consortium. The Chinese author acknowledges support through the Ministry of Science and Technology of the Peoples Republic of China (Grant No. 2018YFA0404803, 2016YFA0401102), the Strategic Priority Research Program of Chinese Academy of Sciences (Grant No. XDB 16010000), the National Science Foundation of China (No. 11875307, 11674339, 11922515, 1193500), the Innovation Program of Shanghai Municipal Education Commission and the Recruitment Program for Young Professionals. We thank Jenny List and  Kristjan Poder (DESY) for contributing section \ref{Electron polarimetry}.


\end{document}